\documentclass[aps,floats,epsf,twocolumn]{revtex4}
 \usepackage{graphics}

\begin{document}

\title{Theory of integer quantum Hall effect in graphene}

\author{Igor F. Herbut}

\affiliation{Department of Physics, Simon Fraser University,
 Burnaby, British Columbia, Canada V5A 1S6}

\begin{abstract} The observed quantization of the Hall conductivity
in graphene at high magnetic fields is explained as being due to the
dynamically generated spatial modulation of either the electron spin
or the density, as decided by the details of Coulomb interaction on
the scale of lattice constant. It is predicted that at a large
in-plane component of the magnetic field such ordering may be
present only at the filling factor $f=\pm 1$ and absent otherwise.
Other experimental consequences of the theory are outlined.
\end{abstract}
\maketitle

\vspace{10pt}

   Graphene is a two-dimensional semi-metal with gapless Dirac-like excitations near
two points in the Brillouin zone \cite{wallace}, \cite{semenoff}.
When placed in the uniform magnetic field of several Tesla it
exhibits plateaus in the Hall conductivity at integer filling
factors $f= 4n+2$, \cite{novoselov}, \cite{zhang}, just as implied
by the Landau level (LL) spectrum of the single-particle Dirac
equation \cite{sharapov}, with each LL being four-fold degenerate in
spin and sublattice indices. Although the strength of the Coulomb
interaction between the conducting electrons in graphene is similar
to the bandwidth, the semi-metallic ground state at zero magnetic
field indicates that it is below the critical value needed for
insulation \cite{khveshchenko}, \cite{herbut}.

At magnetic fields above $\sim 10T$, however, additional quantum
Hall (QH) states at $f=0, \pm 1, \pm 4$ appear \cite{zhang1}.
Plateaus in the Hall conductivity at other integers, such as at
$f=\pm 3$ or $f=\pm 5$ for example, are at the same time
conspicuously absent, even at the highest magnetic field of  $ 45T$.
The experiment \cite{zhang1} also suggests that the activation gaps
at $f=\pm 4$ are likely to be due to the Zeeman splitting of the
first LL. The quantization at $f=\pm 1$, however, implies that the
fourfold degeneracy of the zeroth LL has been completely lifted,
which calls for the inclusion of the Coulomb interactions into
consideration. The theory of the integer QH effect in graphene would
thus have at least to provide answers to the following questions: a)
why has the fourfold degeneracy of {\it only} the $n=0$ LL been
completely lifted at the magnetic fields and samples under study, b)
why do new incompressible states at $f=0,\pm 1, \pm 4$ require
higher magnetic fields to appear than those at $f=4n+2$, and
finally, c) what is the nature of the interacting ground states at
different filling factors?

 The first question may be
immediately answered by postulating that the Dirac fermions have
acquired an effective gap in a form of a `relativistic mass' due to
the Coulomb interaction  \cite{catalysis}. Such a gap reduces the
degeneracy of {\it only} the zeroth LL. Including the Zeeman
splitting then leads to the spectrum of the effective
single-particle Hamiltonian precisely as required by the observed
pattern of quantization of Hall conductivity. It is unclear,
however, under which circumstances, and which one of the multitude
of such gaps that could exist, as discussed below,  is generated
dynamically. This problem is addressed here within the extended
Hubbard model on a honeycomb lattice, with both on-site and
nearest-neighbor repulsions. This is the simplest Hamiltonian that
mimics the effect of Coulomb repulsion and which contains the
possibilities of charge-density-wave (CDW) and antiferromagnetic
(AF) orders. Previous work indicated \cite{herbut},
\cite{tchougreeff} that these two are in direct competition at zero
magnetic field and when the interaction is strong. Here we solve the
model in the physical {\it weak} coupling regime and in an external
magnetic field, allowing only for the simplest AF state with the
staggered magnetization parallel to the magnetic field. The results
are summarized as the phase diagrams in Figures 1 and 2. At
half-filling and for a weak enough Zeeman energy the system could be
either a CDW or an AF, depending on which coupling dominates (Fig.
2). For a larger Zeeman energy the ground state at $f=0$ becomes
magnetic with full lattice symmetry. Nevertheless, even in the
latter case increasing the chemical potential produces an
incompressible state at $f=1$, with the activation gap becoming
equal to the `relativistic' gap (Fig. 1). In contrast, at weak
Zeeman coupling we find a direct transition between $f=0$ and $f=2$
states. At $f\geq 2$ the `relativistic' gap vanishes. Experiments
that would test the presented against other theories \cite{gusynin},
\cite{alicea}, \cite{fuchs}, \cite{nomura} are discussed.

\begin{figure}[t]
{\centering\resizebox*{80mm}{!}{\includegraphics{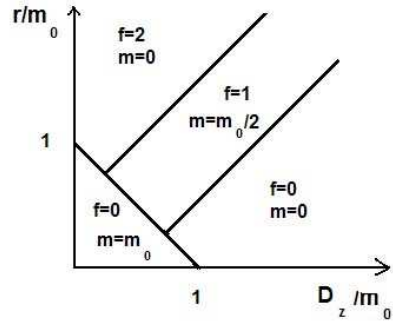}}
\par} \caption[] {The proposed phase diagram of graphene in the
magnetic field. $D_z= g_z B$ is the Zeeman energy, $r$ is the
chemical potential, and $m_0 =2 g_x B_\bot /\pi$ is the
characteristic size of the `relativistic' many-body gap $m$. $g_x$
is the larger of the couplings in the CDW and AF channels (see the
text).}
\end{figure}

We define the extended Hubbard model as $H=H_0 + H_1$, where
\begin{equation}
H_0= -t \sum_{\vec{A}, i, \sigma=\pm} u^\dagger _\sigma (\vec{A})
v_\sigma (\vec{A}+\vec{b}_i) + H. c.,
\end{equation}
\begin{equation}
H_1 = U \sum_{\vec{X}} n_+(\vec{X}) n_{-} (\vec{X}) + \frac{V}{2}
\sum_{\vec{A},i, \sigma, \sigma'} n_\sigma (\vec{A}) n_{\sigma'}
(\vec{A}+\vec{b}_i).
\end{equation}
The sites $\vec{A}$ denote one triangular sublattice of the
honeycomb lattice, generated by linear combinations of the basis
vectors $\vec{a}_1= (\sqrt{3}, -1)(a/2)$ and $\vec{a}_2 = (0,a)$.
The second sublattice is then at $\vec{B}=\vec{A}+\vec{b}$, with the
vector $\vec{b}$ being either $\vec{b}_1= (1/\sqrt{3},1) (a/2)$,
$\vec{b}_2= (1/\sqrt{3},-1) (a/2)$, or $\vec{b}_3= (-a/\sqrt{3},0)$.
$a\approx 2.5 A$ is the lattice spacing, and $t\approx 2.5 eV$,
$U\approx 5-12 eV$, and $U/V \approx 2-3$ \cite{tchougreeff}.

\begin{figure}[t]
{\centering\resizebox*{80mm}{!}{\includegraphics{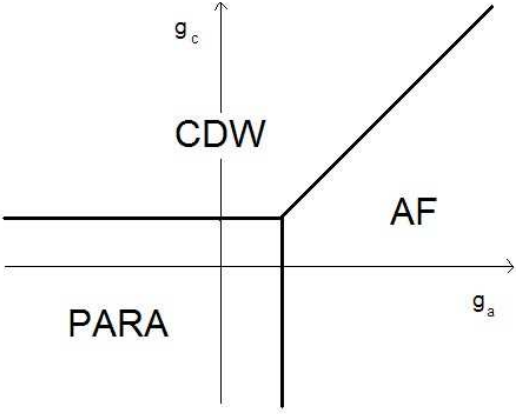}}
\par} \caption[] {The phase diagram at half-filling ($r=0$). The
translationally symmetric magnet exists for $g_c, g_a < \pi g_z
B/2B_\bot$.}
\end{figure}

 The spectrum of $H_0$ becomes linear in the vicinity of the two Dirac
points at $\pm \vec{K}$, with $\vec{K} = (1,1/\sqrt{3}) (2\pi/a
\sqrt{3})$ \cite{wallace}, \cite{semenoff}. Retaining only the
Fourier components near $\pm \vec{K}$ one can write, in the
continuum notation,
\begin{equation}
H_0= \int d\vec{x} \sum_{\sigma} \Psi^\dagger _\sigma (\vec{x}) i
\gamma_0 \gamma_i D_i \Psi_\sigma (\vec{x}),
\end{equation}
and
\begin{eqnarray}
\Psi_\sigma ^\dagger (\vec{x}) = \int^\Lambda \frac{d\vec{q}}{(2\pi
a )^2} e^{ -i \vec{q}\cdot \vec{x}} (u^\dagger _\sigma
(\vec{K}+\vec{q}), \\ \nonumber
 v^\dagger _\sigma (\vec{K}+\vec{q}),
 u^\dagger _\sigma(-\vec{K}+\vec{q}), v^\dagger _\sigma
(-\vec{K}+\vec{q}) ),
\end{eqnarray}
$i=1,2$, where it was convenient to rotate the reference frame so
that $q_1 = \vec{q}\cdot \vec{K}/K$ and $q_2 = (\vec{K}\times
\vec{q})\times \vec{K}/K^2 $, and set $\hbar=e/c=v_F =1$, where $v_F
=ta\sqrt{3}/2$ is the Fermi velocity. Here $i\gamma_0 \gamma_1 =
\sigma_z \otimes \sigma_x$, $i \gamma_0 \gamma_2 = - I_2 \otimes
\sigma_y$, with $I_2$ as the $2 \times 2$ unit matrix, and
$\vec{\sigma}$ as the Pauli matrices. $\Lambda\approx 1/a $ is the
ultraviolet cutoff over which the linear approximation for the
dispersion holds. The orbital effect of the magnetic field is
included by defining $D_i = -i\partial_i - A_i$, with its component
perpendicular to the graphene's  plane being $B_\bot = \partial_1
A_2 -
\partial_2 A_1$.

 Consider an auxiliary  single-particle Hamiltonian $\tilde{H}$:
\begin{equation}
\tilde{H} =  m M + i \gamma_0 \gamma_i D_i,
\end{equation}
where $M$ is a Hermitian $4\times 4$ matrix.  When $m=0$,
$\tilde{H}=H_0 $, for each spin state. Also, if $M^2 -1 =\{ M,
\gamma_0 \gamma_i \} =0$,
\begin{equation}
\tilde{H}^2 =  D_{i} ^2 + B (\sigma_z \otimes \sigma_z) + m^2.
\end{equation}
 This is the case if either
\begin{equation}
M= M_1= a (I_2 \otimes \sigma_z ) + b (\sigma_x \otimes \sigma_x) +c
(\sigma_y \otimes \sigma_x)
\end{equation}
with real $a,b,c$ which satisfy  $a^2 + b^2 + c^2 =1$, or
\begin{equation}
M= M_2= \sigma_z \otimes \sigma_z.
\end{equation}
In either case the eigenvalues  of $\tilde{H}^2$ are at $2nB+ m^2$,
with $n=0,1,2,...$. For $n>0$ this immediately implies that
eigenvalues of $\tilde{H}$ itself are  at $\pm \sqrt{2nB+ m^2}$,
with the degeneracies of $B/\pi$ per unit area being the same as for
$m=0$. For $n=0$, on the other hand, an elementary calculation gives
that the eigenvalues of $\tilde{H}$, a) for any $M_1$, are at $\pm
|m|$, each with halved degeneracy of $B/2\pi$ (per unit area), and,
b) for $M_2$, are at $m$, still with the full degeneracy of $B/\pi$.
The invariance of the spectrum in the first case under rotations of
the unit vector $(a,b,c)$ is the consequence of the `chiral' $SU(2)$
symmetry of $H_0$ generated by $\{ \gamma_{35}, \gamma_3,
\gamma_5\}$, where $\gamma_3 = \sigma_x \otimes \sigma_y$, $\gamma_5
= \sigma_y \otimes \sigma_y$, and $\gamma_{35} = i\gamma_3
\gamma_5$, for example \cite{herbut}, \cite{herbut1}. Any specific
choice of $M_1$ in $\tilde{H}$ breaks this $SU(2)$ down to $U(1)$,
and leads to the same eigenvalues. $M_2$ preserves the chiral
symmetry, and hence implies a different spectrum.

Assuming that such an effective chiral-symmetry-breaking gap $m$
becomes generated by interactions and then splitting the resulting
energy levels further by the standard Zeeman effect could thus lead
to the QH states at all even $f$, $f=0$, and $f=\pm 1$, in
accordance with experiment \cite{zhang1}. Here the filling factor is
$f=2\pi N/B$, with $N$ as the number of electrons measured from a
half-filled band. This still, however, leaves a complete freedom of
choice of the vector $(a,b,c)$. In particular, this choice may
differ for the two spin states as well. For example, at $B=0$ and at
strong $V$, one expects the interactions to prefer $a=1$ for both
projections of spin, i. e. the CDW \cite{semenoff}. For strong $U$
and at $B=0$, on the other hand, it is $a=\sigma$, i. e. the AF,
that has lower energy \cite{herbut}. $a=0$ would correspond to the
competing `Kekule' ordering \cite{hou}. Which direction on the
chiral manifold is actually chosen by the system is thus obviously a
question of dynamics, to which we turn next.

The low energy Lagrangian for the extended Hubbard model may be
written as \cite{herbut}
\begin{eqnarray}
L= i\sum_\sigma \bar{\Psi}_\sigma \gamma_\mu D_\mu \Psi_\sigma -
\sum_\sigma (r +\sigma g_z B) \Psi^\dagger _\sigma \Psi_\sigma \\
-\nonumber g_c (\sum_{\sigma } \bar{\Psi}_\sigma \Psi_\sigma )^2 -
g_a (\sum_{\sigma } \sigma \bar{\Psi}_\sigma \Psi_\sigma)^2,
\end{eqnarray}
where $\bar{\Psi}_\sigma =  \Psi^\dagger _\sigma (\vec{x},\tau)
\gamma_0$, $D_0 =-i \partial_\tau$, with $\tau$ as the imaginary
time, $\mu=0,1,2$, and $\gamma_0= I_2 \otimes \sigma_z$, $\gamma_1=
\sigma_z \otimes \sigma_y$, and $\gamma_2= I_2 \otimes \sigma_x$.
Here $g_a\approx U a^2/8$, $g_c\approx (3 V-U)a^2/8$, $g_z$ is the
(dimensionless) effective $g$-factor of the electron, and $B=(B_\bot
^2 + B_ \| ^2 )^{1/2} $, with $B_\|$ as the field's in-plane
component. We have retained only the {\it two least irrelevant}
short-range interactions among those present in the full effective
Lagrangian at $B=0$ \cite{herbut}. This will prove justified at the
magnetic fields of interest, as discussed shortly. We have also set
the mass of the electron $m_e$ to unity.

Performing the Hubbard-Stratonovich transformation and neglecting
the quantum fluctuations the free energy per unit area and at $T=0$
may be written as
\begin{eqnarray}
F(m_c, m_a) - F(0, 0) =  \frac{m_c ^2}{4g_c}+ \frac{m_a ^2}{4g_a} + \\
\nonumber \frac{B_\bot}{4\pi^{3/2} } \int_{0} ^\infty
\frac{ds}{s^{3/2} } \sum_{\sigma=\pm} ( e^{-s m_\sigma ^2}
-1)f_\sigma (s, m_\sigma ),
\end{eqnarray}
where $m_\sigma =  m_c +\sigma m_a$, and the function
\begin{eqnarray}
f_\sigma (s,m) = \theta ( |m|- |r + \sigma g_z B| )+ \\
\nonumber C(s\Lambda^2) ( \coth(sB_\bot)-1).
\end{eqnarray}
$m_c/g_c  = \sum_\sigma \langle \bar{\Psi}_\sigma \Psi_\sigma
\rangle$, $m_a/g_a = \sum_\sigma \sigma \langle \bar{\Psi}_\sigma
\Psi_\sigma \rangle$,  are the CDW and the AF order parameters
\cite{herbut}, respectively. $C(x)$ is the cutoff function that
satisfies $C(x \rightarrow \infty) = 1$ and $C(x\rightarrow 0) = 0$.
Summing over the LLs below a sharp cutoff in energy, for example,
yields $C(x)=1-e^{-x}$. The first term in Eq. (11) represents the
crucial zeroth LL contribution, and the second the remaining LLs.
Zeeman energy is taken to always be smaller than the separation
between the zeroth and the first LL, i. e. $g_z B < \sqrt{2
B_\bot}$.

Let us first consider the system at $r=0$. The free energy is
minimized by the solution of
\begin{equation}
\frac{m_+}{g_+} + \frac{m_-}{g_-} = \frac{4B_\bot m_+}{\pi^{3/2}}
\int_0 ^\infty \frac{ds}{s^{1/2}} e^{-s m_+ ^2 } f_+ (s, m_+),
\end{equation}
\begin{equation}
\frac{m_-}{g_+} + \frac{m_+}{g_-} = \frac{4B_\bot m_- }{\pi^{3/2}}
\int_0 ^\infty \frac{ds}{s^{1/2}} e^{-s m_- ^2 } f_- (s, m_- ),
\end{equation}
where $g_\pm ^{-1} = g_c ^{-1} \pm g_a ^{-1}$. Assume both $g_c$ and
$g_a$ to be weak, $\Lambda g_{c,a} \ll 1$, and positive, and $g_c
> g_a$, for example. There are then three types of solutions:

 1) If $m_+ \geq m_- > g_z B$, then  $m_+ = m_-$, i. e.
$m_a=0$ and $m_c = 2g_c B_\bot /\pi$.  This is the CDW. It exists
when $g_c > \pi g_z B/2B_\bot$. For $g_a
> g_c$, of course, one finds $m_c=0$ and $m_a= 2g_a B_\bot/\pi$, i. e.
the AF (Fig. 2). This, in particular, includes the case of the pure
Hubbard model with $V=0$. The linear dependence on $B_\bot$ reflects
the proportionality of either $m$ to the degeneracy of the LLs.

2) For $g_z B > m_+ \geq m_- $ one finds a paramagnet with $m_c =
m_a = 0$, unless the stronger coupling exceeds the $B=0$ critical
value of $\pi/8\Lambda$, which lies outside the weak-coupling
regime.

3) Finally, when $m_+ > g_z B > |m_-| $, the solution is $m_x = g_x
B_\bot / \pi$, where $x=c,a$. Both the CDW and AF order parameters
are finite, which would corresponds to a {\it ferrimagnet}. This
solution exists for $g_c + g_a > \pi g_z B/ B_\bot$, and $|g_c -
g_a| < \pi g_z B/ B_\bot$.

  At weak coupling only the first term, representing the
 zeroth LL in Eq. (11), actually matters for the gap $m$. The second term would become important only at
 strong couplings, or in the limit $B_\bot \rightarrow 0$. Comparing
 the energies of the four possible solutions, we find only three stable phases represented in Fig. 2.
 All three have the filling factor $f=0$.
 Even if both $g_c$ and $g_a$ are positive it is always energetically favorable to open a single
but larger gap that corresponds to the dominant coupling. The
aforementioned ferrimagnet, which would have $f=1$,  is thus never
the minimum of the energy at $r=0$. This may also be understood as
follows. At low magnetic fields, such that $l_B \gg a$ where $l_B =
1/\sqrt{B_\bot}$ is the magnetic length in our units, the flow of
the couplings implied by the invariance of the gaps in Eqs.
(12)-(13) with respect to change of the cutoff $\Lambda$, in the
regime $\Lambda \gg 1/l_B$ is the same as at $B_\bot =0$
\cite{kaveh}. All weak couplings are thus irrelevant at intermediate
length scales between the lattice constant and the magnetic length
\cite{herbut}. At $B_\bot \neq 0$ the flow of the interaction
couplings towards the stable Gaussian fixed point is terminated at
the cutoff $\sim 1/l_B$, with the least irrelevant coupling being
left as dominant in the low energy theory. This single surviving
coupling then selects and generates the `relativistic' gap at
$B_\bot \neq 0$. All other more irrelevant couplings can be
neglected, as we partially did in Eq. (9), from the outset.
Incidentally, this also justifies the Hartree approximation to the
free energy utilized in Eq. (10): as long as the ground state is
semimetallic and gapless at $B_\bot =0$, for a weak enough magnetic
field the low energy theory at the  cutoff $\sim 1/l_B \ll 1/a$ is
indeed weakly interacting, as assumed.

It is useful to display explicitly the relevant energy scales in
   the problem. First, the `relativistic' gap
   is $m\approx (U/8) (B_\bot / B_0)$, where $B_0=1/a^2 \approx 10^5 T$ is the
   characteristic scale for the magnetic field set by the lattice constant.
    Assuming $U\approx 10 e V$
   gives  $m\approx 1 meV$ for $B_\bot\sim 10 T$.
   The LL separation, on the other
   hand, is $D_L \approx t (B_\bot /B_0)^{1/2}$, and thus
   by roughly two orders of magnitude larger. The
   Zeeman energy, on the other hand, for $g_z \approx 1$
   is $D_z \approx (B/10^4 T) eV$, and thus of a similar size as $m$. Since the widths of the
plateaus at $f=0, \pm 1, \pm 4$ are proportional to either $m$ or
$D_z$, both much smaller than  $D_L $, this would naturally explain
why these QH states require higher magnetic fields to become
discernible in presence of some fixed disorder-induced broadening of
the LLs.

 The above discrepancy between the kinetic and the interaction energy
scales is only the first in a general hierarchy of energy scales
that is implied by the stability of the semimetallic  fixed point at
zero magnetic field. As discussed above, the magnetic field affects
the flow of the interaction couplings only at the length scales
above the magnetic length $l$. The development of an order parameter
$m_x$ in some channel may be understood as the divergence of the
corresponding coupling constant $x$ at the length scale $L_x = 1/m_x
\gg l$. This is determined by $x$, via essentially dimensional
analysis, by the relation $m_x l \approx x l^{-n}$, where $x l^{-n}$
is the dimensionless coupling, with $1/l$ now playing the role of
the ultraviolet cutoff. For $x=g$ and $n=1$ this way we reproduce
the above relation for the interaction gap $m$, and for $x=t$ and
$n=0$ we get the characteristic kinetic energy scale $D_L$. The
crucial point is that more irrelevant interactions at $B=0$, with
higher negative dimensions $n$, translate at $B\neq 0$ into energy
scales that depend on higher powers of $B/B_0$: $m_x \sim
(B/B)^{(n+1) /2}$. In particular, this is what justifies the
replacement of a realistic finite-range interaction with the simpler
$\delta$-function in Eq. (9).

  At $r=0$ there are thus two fundamentally different ground states: one at
  larger interactions
  that breaks the $A-B$ sublattice symmetry, either in the charge (CDW) or the spin
  (AF) channel, with vanishing magnetization, and the other magnetic,
  at weaker interactions, with the full translational symmetry of the
  lattice. Both are incompressible
  and yield the plateau in  the Hall conductivity at
  $f=0$. If the parameters are such to place the
  system for $B_\| =0$  into the former, increasing $B_\|$ would
   eventually cause the transition into the latter state. This certainly holds
   for the CDW, as well as for the AF with its Neel order directed along
   the magnetic field, as we assumed here \cite{comm}.

Hereafter we will thus retain only the larger quartic coupling, call
it $g_x$, and its corresponding order parameter in Eq. (10), $m$. At
small $m$ the free energy then becomes
\begin{eqnarray}
\Delta F= \frac{m^2}{4 g_x} (1+ O(\Lambda g_x))  + O(m^4) \\
\nonumber - \frac{B_\bot |m|}{2 \pi} (\theta (|m|- |r+ D_z|)+
\theta(|m|-|r-D_z|))
\end{eqnarray}
where $D_z=g_z B$. The non-analytic term $\sim |m|$ comes from the
zeroth LL. In the weak-coupling regime possible local minima are
then at $m/m_0 = 0,1/2,1$, with $m_0= 2 g_x B_\bot/\pi$.
Determination of the global minimum of the free energy at a finite
chemical potential then leads to the phase diagram presented in Fig.
1. Let us determine the filling factor when $m=m_0/2$ solution is
stable, i. e. in the strip  $r+D_z > m_0$ and $|r-D_z|< m_0 /2$ in
Fig. 1. From the first inequality we see that $r+D_z
> m$, and thus $r>m-D_z>-m-D_z$, so the values of two
out of four spin- and sublattice-degeneracy-resolved energies of the
zeroth LLs lie below the chemical potential. The second inequality
implies that between the two remaining energies one is above and the
other below the chemical potential. To see this, first assume
$r>D_z$. Then $D_z - m < r < D_z +m$, as announced. If $r<D_z$, on
the other hand, $D_z + m
> r > D_z -m$ again, so in either case $f=1$. The determination of
the filling factor for the remaining states is even simpler and
follows analogously.

For $f\geq 2$ the `relativistic' gap is $m=0$, and there is a direct
transition between $f=0$ to $f=2$ at $D_z / m_0 <1/4$. For a fixed
and larger $D_z$, by increasing the chemical potential the system
always passes through the intermediate $f=1$ QH state, which is
magnetic and breaks the discrete sublattice symmetry. The width of
$f=1$ state is $2D_z - (m_0 / 2)$ for $D_z/m_0 < 3/4$, or $m_0$ for
$ D_z /m_0 > 3/4$. All transitions are discontinuous.
 As the ratio $ D_z /m_0$ can be changed by varying $B_\|$, it follows that at a
large enough $B_\|$ the width and the activation energy of the
$f=\pm 1$ state becomes $\sim B_\bot$ and independent of $B_\|$,
which should be experimentally testable. Since in the experiment
\cite{zhang1} the width of the plateau at $f=0$ appears to be
somewhat larger than at $f=\pm 1$ and $f=\pm 4$, the latter being
always $2D_z$, we speculate that $1/4 < D_z /m_0 < 1/2$.
 If this is indeed the case the activation energy at $f=0$ should first
decrease before increasing with $B_\|$, whereas that of $f=1$ would
increase and saturate.

  The mechanism of `magnetic catalysis' of the `relativistic' gap \cite{catalysis}
  utilized here has also been considered recently as an explanation of the integer QH effect in graphene
  in ref. \cite{gusynin}. The crucial
  difference from the present work is that only $\sim 1/r$ tail of
  the  Coulomb interaction was included, which could produce only the CDW, and
  that with the large  gap $m\sim \sqrt{B_\bot }\sim D_L$.
  To have $m$ much smaller from the LL separation, as observed experimentally, requires then what appears to be
  an unrealistically strong screening of the  Coulomb interaction by the substrate \cite{gusynin}.
  In this theory the gap is also always inhibited
  by a finite chemical potential, and thus cannot exist at $f=1$
  if it did not already at $f=0$. This would imply  that the activation gap
  in \cite{gusynin} at $f=1$ is always the Zeeman energy. Both
  results are in sharp contradiction with ours.

 The QH effect in graphene was also recently discussed in \cite{alicea}.
 The principal difference from the present work is that the AF ordering was entirely neglected, so that the
 existence of the QH state at $f=1$ in \cite{alicea} is always due to the CDW,
 which then requires a large enough $V$. In contrast, in the present theory
 $f=1$ state can exist even for $V=0$, when it is due to the AF ordering. In particular,
 setting $V=g_z =0$ we find the pure ($V=0$) weakly coupled Hubbard model at half filling and in magnetic field
 to be an AF, and not the Stoner's ferromagnet, as assumed in \cite{alicea}.
 This is because a finite AF order parameter, unlike the magnetization, besides
splitting the zeroth LL, also lowers the energy of all other
occupied LLs. Put differently, in considering the dynamics within
only the zeroth LL, the couplings need to be renormalized up to the
length scale of the magnetic length $l_B$ first. Then $g_a (l_B )
\gg g_f (l_B )$ always, where $g_f$ is the coupling in the
ferromagnetic channel \cite{herbut}, for this reason omitted in Eq.
(9). Retaining such a ferromagnetic coupling may be shown to only
increase the activation energy when $m=0$ to $D_z + 2 g_f B_\bot
/\pi$.

CDW formation due to lattice distortion has been proposed as a
mechanism behind the QH effect in graphene in \cite{fuchs}. This
explanation differs from the present and all the other ones in that
Coulomb repulsion plays essentially no role in it. In particular, an
$f=1$ incompressible state would in that scenario appear with
increase of the chemical potential for any Zeeman energy, in
contrast to our Fig. 1, where for weak Zeeman energy there is a
direct transition from $f=0$ to $f=2$.

  Finally, the present mechanism differs essentially from the recent
  proposal \cite{nomura}, in which disorder is invoked
  to explain the absence of QH states at odd $f\neq \pm 1$.

 In this paper we have neglected entirely the effect of disorder.
 As usual, it will broaden the LLs, and thus provide an intrinsic
 energy scale that needs to be exceeded in order for an
 incompressible state to be resolved. We believe that this is why the states at
 $f=0$, $f=\pm 1$, and $f=\pm 4$ become visible only at higher
 fields than the main sequence at $f=4n+2$, since their energy gaps
 according to our scenario are inherently smaller. They may be
 therefore understood as the {\it fine structure} of the QH effect
 in graphene.

   To summarize, we assumed that the principal effect of the Coulomb
 interaction in graphene is to introduce the on-site and nearest-neighbor
 repulsion for electrons on a honeycomb lattice. Postulating further a
 semi-metallic ground state in zero field, we argued that at a finite field there
 exists a hierarchy of energy scales, determined by the degrees of
 irrelevancy of the corresponding couplings at $B=0$. In this work
 we considered only the effects of the least irrelevant interaction,
 which we expect to give the leading correction to
 the Landau level structure for the non-interacting electrons.
The phase diagram of graphene at
 laboratory magnetic fields $\sim 10T$ is proposed. The theory predicts
 the incompressible states at all even integer filling factors including
 $f=0$, and at $f=\pm 1$. The
 ground state of the system at $f=0$ either breaks the sublattice
  symmetry, in either charge or spin channels, or is magnetic,
 depending on the magnitude of the Zeeman energy.
 At $f=\pm 1$ the system is always in the translational-symmetry-breaking
 phase and with finite magnetization, whereas at $|f| \geq 2$ the
 sublattice symmetry is preserved. The phase diagram in the Zeeman
 energy - chemical potential plane is proposed, and several of its
 features that may be used to distinguish the present from other
 proposed scenarios are discussed.

The author is grateful to J. Alicea, J.-N. Fuchs. V. Gusynin, and S.
Sharapov for useful discussions and correspondence.  This work was
supported by the NSERC of Canada.

\end{document}